# Thermodynamic perturbation theory for associating fluids confined in a 1-dimensional pore

Bennett D. Marshall

*ExxonMobil Research and Engineering, 22777 Springwoods Village Parkway, Spring TX 77384*

## Abstract

In this paper a new theory is developed for the self – assembly of associating molecules confined to a single spatial dimension, but allowed to explore all orientation angles. The interplay of the anisotropy of the pair potential and the low dimensional space, results in orientationally ordered associated clusters. This local order enhances association due to a decrease in orientational entropy. Unlike bulk 3D fluids which are orientationally homogeneous, association in 1D necessitates the self – consistent calculation of the orientational distribution function. To test the new theory, Monte Carlo simulations are performed and the theory is found to be accurate. It is also shown that the traditional treatment in first order perturbation theory fails to accurately describe this system. The theory developed in this paper may be used as a tool to study hydrogen bonding of molecules in 1D zeolites as well as the hydrogen bonding of molecules in carbon nanotubes.

# I. Introduction

Associating molecules have short ranged highly directional intermolecular potentials resulting in a limited valence of the association interaction. The hydrogen bond is the most common example, however more recent examples include patchy colloids[1-3] and globular proteins[4]. A successful theoretical formalism to describe associating fluids is the multi – density formalism of Wertheim.[5-7] In this approach each bonding state of an associating molecule is assigned a separate density, this allows for the incorporation of steric effects at an early point in theory development, ultimately reproducing the limited valence of the association bond even when approximations of the theory are used, such as thermodynamic perturbation theory (TPT). The approach is typically applied at the level of first order perturbation theory (TPT1) where each association bond is treated independently and is assumed singly bondable (monovalent). In addition, TPT1 typically assumes that the density is not a function of orientation meaning all orientations are equally probable. This allows for a simple form of the Helmholtz free energy for molecules with any number of association sites. [8, 9] TPT1 provides the basis for the popular SAFT[10-12] equation state, and has also been extensively applied in the modelling of patchy colloid[13-15] fluids.

In recent years there has been interest in expanding the applicability of TPT beyond the constraints of TPT1. For instance, TPT has now been extended to include the effect of bond angle on association[16, 17], bond cooperativity[18, 19] as well as the possibility of forming multiple bonds per association site.[20-29]

Another area where a traditional application of TPT1 will prove inadequate is that of associating fluids in low dimensional spaces (1 or 2 spatial dimensions), which can explore their full three dimensional orientation space. For these cases, the fluid will no longer be orientationally homogeneous as in 3D bulk fluids. The anisotropy of the pair potential, combined with low dimensional space, will induce local orientational order within clusters. This results in a non – constant orientational distribution function (ODF). That is, once a molecule is bonded its orientation is restricted to be in a certain range with respect to the spatial axis. This decrease in orientational entropy increases the probability that a bonded molecule will be oriented to bond to a second molecule.

Density functional theories for inhomogeneous associating fluids based on TPT1[30-33] have been applied to confined associating systems. However, the theories will not have the correct narrow pore 1D limit, as the association is treated in TPT1 with no inclusion of the decrease in orientation states resulting from association in 1D. TPT1 has been applied to 2D[34,35] systems where the association sites are restricted to being parallel to the substrate; however, this approach will not capture the orientation dependence of association. In addition, Kierlik and Rosinberg (KR)[36] demonstrated the applicability of TPT to the case of 1D associating fluids. However, the treatment of KR was for a simple spin representation of orientation with no path forward to calculate the effect of association on the system ODF. We show here, that in order to accurately model low dimensional associating systems, the effect of association on the ODF must be included in any theoretical treatment.

In this work we extend TPT to the case of a two site associating fluid confined to a single spatial dimension, yet allowed to explore all orientation angles. The development of the association contribution of the theory is more complex than the bulk 3D case due to the fact that the system is orientationally inhomogeneous, and we must solve for the ODF. It is found that in these 1D systems association is intimately related to order in clusters. To validate the theory we perform new Monte Carlo simulations and the theory is found to be accurate. Far from being of solely theoretical interest, this new theory could be used to describe 1D hydrogen bonded wires in zeolites[37-40] and carbon nanotubes[41, 42] where it has been shown that water hydrogen bonding dominates the fluid structure, leading to long ranged orientational order. Lattice models[43, 44] have been used to describe these systems thus far, the approach developed in this paper will provide a continuum alternative to these lattice methods.

## II. Theory

In this section we develop the new theory for associating fluids confined to a single spatial dimension $z$. In addition to this single spatial degree of freedom we allow the molecules to freely explore all orientation angles. For the intermolecular potential considered in this work we need only consider a single orientation angle $\theta$ with respect to the axis $z$. We consider the case of a fluid of hard spheres of diameter $d$ decorated with two conical square well (CSW) association sites of differing functionality located on opposite poles of the sphere as pictured in Fig. 1.

The intermolecular potential for molecules of this type is given as

$$\phi(12) = \phi_{HS}(z_{12}) - \varepsilon_{AB}(O_{AB}(12) + O_{BA}(12)) \tag{1}$$

where $(1) = (z_1, \theta_1)$ is short hand for the position and orientation of sphere 1. The potential $\phi_{HS}(z_{12})$ is the hard sphere reference potential, which in 1D is given as

$$\phi_{HS}(z_{12}) = \begin{cases} \infty & z_{12} < d \\ 0 & otherwise \end{cases} \tag{2}$$

where $z_{12} = |z_2 - z_1|$ is the distance between the centers of two spheres.

In Eq. (1) we have restricted association to be between unlike sites; that is there are $AB$ bonds, but no $AA$ or $BB$ bonds. The quantity $O_{AB}$ is the overlap function given by

$$O_{AB}(12) = \begin{cases} 1 & z_{12} \leq z_c \text{ and } \theta_{A1} \leq \theta_c \text{ and } \theta_{B2} \leq \theta_c \\ 0 & otherwise \end{cases} \tag{3}$$

Where $z_c$ is the maximum distance between spheres for which association can occur, $\theta_{A1}$ is the angle between the center of site $A$ on sphere 1 and the coordinate $z$, and $\theta_c$ is the maximum angle for which association can occur. With this, if two spheres are both positioned and oriented correctly, a bond is formed and the energy of the system is decreased by a factor $\varepsilon_{AB}$. CSW sites have been widely used as both primitive models for hydrogen bonding fluids as well as patchy colloids. [15, 27, 45] Figure 2 outlines a 1D fluid of associating spheres. Note, $\theta$ is the polar angle in a spherical coordinate system defining the orientation vector. There is also an azimuthal angle $0 \leq \varphi \leq 2\pi$; however, with this spherical coordinate system, this angle does not affect Eq. (1).

To develop the free energy we employ Wertheim's multi-density formalism[5-7] for associating fluids. We consider the fluid to exist in a single dimension of length $L$ and in the absence of any external fields. In bulk fluids in three dimensions, in addition to translational homogeneity, there is also rotational homogeneity. Meaning there is no preferred orientation of the molecules. This is not the case here, due to the anisotropic pair potential in Eq. (1) combined with the single spatial dimension. That is, for spheres to associate they must be oriented where the $z$ axis passes through the CSW association sites. The system is not orientationally homogeneous and the density of spheres in the fluid will, in general, be a function of orientation $\rho(\theta)$. In this sense, somewhat surprisingly, the 1D case is more challenging than the 3D case where the density is independent of orientation.

Wertheim's multi – density formalism is typically applied as a perturbation theory (TPT) to a hard sphere reference fluid. TPT is derived by neglecting all graphs to the cluster sum which contain more than a single path of association bonds. This allows the theory to be written in terms of correlation functions of the reference fluid only. The remaining graphs are then ordered by the number of association bonds in the graphs with first order perturbation theory (TPT1) retaining all diagrams with up to one association bond, TPT2 retains all diagrams with up to two association bonds etc…. As discussed by Kierlik and Rosinberg (KR)[46] in 1D all contributions for TPT2 and higher cancel due to the fact that the linear superposition of the *n*-body correlation function holds exactly in 1-D.

Here we start with the Helmholtz free energy[6, 27] for two site associating fluids, allowing for orientational inhomogeneities in the ideal and association contributions

$$\frac{A - A_{HS}^{EX}}{k_b TL} = \int_{-1}^{1} db \rho(b) \left( \ln(\rho(b)\Lambda) + \sum_{C=A,B} \left( \ln X_C(b) - \frac{X_C(b)}{2} \right) \right) \tag{4}$$

where $b = cos(\theta)$. Expression (4) is a functional due to the fact that density is a function of orientation. The free energy $A_{HS}^{EX}$ is the excess free energy of a system of 1D hard spheres which is known exactly[47]

$$\frac{A_{HS}^{EX}}{Nk_b T} = \ln \frac{1}{1 - \bar{\rho}d} \tag{5}$$

Where *N* is the number of spheres and

$$\bar{\rho} = \int_{-1}^{1} \rho(b) db \tag{6}$$

The fractions $X_A$ are the fraction of spheres *not* bonded at site *A*. For the CSW potential in 1D we simplify this fraction as

$$\frac{1}{X_A(b_1)} = 1 + I_A(b_1) \tag{7}$$

The integral $I_A$ is given by (for $L \to \infty$)

$$I_A(b_1) = f_{AB} \int_{-1}^{1} \int_{0}^{\infty} db_2 \rho(b_2) dz_{12} g_{HS}(z_{12}) O_{AB}(12) X_B(b_2) \tag{8}$$

The Mayer function $f_{AB} = \exp(\varepsilon_{AB}/k_B T) - 1$ and $g_{HS}(z_{12})$ is the pair correlation function of the hard sphere reference fluid which in the range $d \leq z_{12} \leq 2d$ is given by[48]

$$g(z_{12}) = \frac{\exp(-\bar{\rho}(z_{12}-d)/(1-\bar{\rho}d))}{1-\bar{\rho}d} \quad (9)$$

Similar equations exist for $X_B$ and are obtained from Eqns. (7) and (8) by interchanging the site labels $A$ and $B$.

Since the density is orientation dependent it will be beneficial to define an orientational distribution function (ODF) as

$$\xi(b) = \frac{\rho(b)}{\bar{\rho}} \quad (10)$$

Combining (6) and (10) find

$$\int_{-1}^{1} \xi(b)db = 1 \quad (11)$$

For the 1D CSW case considered here, we can split the ODF into contributions from unbonded spheres $\xi_o$ and spheres which are bonded $\xi_b$ as

$$\xi(b) = \bar{X}_o \xi_o(b) + (1-\bar{X}_o)\xi_b(b) \quad (12)$$

The orientational average of any quantity $A$ is given by

$$\bar{A} = \int_{-1}^{1} \xi(b)A(b)db \quad (13)$$

The monomer fraction is given by the relation[27]

$$X_o(b) = X_A(b)X_B(b) \quad (14)$$

Combining (13) – (14) we obtain a relation for the average monomer fraction

$$\bar{X}_o = \frac{N_o}{N} = \int_{-1}^{1} \xi(b)X_A(b)X_B(b)db \quad (15)$$

Since there is no external field, the orientation of monomers is unperturbed with an ODF

$$\xi_o(b) = \frac{1}{2} \tag{16}$$

For a sphere to be bonded, it must be in an orientation $\theta \leq \theta_c$ or $\theta \geq \pi - \theta_c$. In the absence of an external field these two cases will occur with equal probability yielding the following ODF for bonded spheres

$$\xi_b(b) = \frac{1}{2} \frac{U(|b| - \cos\theta_c)}{1 - \cos\theta_c} \tag{17}$$

$U(x)$ is the Heaviside step function. Note, Eq. (17) gives an equal probability to the orientations $\theta \leq \theta_c$ or $\theta \geq \pi - \theta_c$. What this means is that to be in a bonded cluster the orientation is restricted to values $|b| \geq \cos\theta_c$, so in this sense there is order in the system as the number of orientation states have been decreased. However, there is no true global order as the ODF (17) assigns equal probabilities to each possible orientation. That is $b < 0$ and $b > 0$ occur with equal probability meaning left handed ($A$ sites pointing left) and right handed clusters ($A$ sites pointing right) occur with equal probability. Summing over all clusters in the fluid there will be no overall order.

Now we can write the overall ODF as

$$\xi(b) = \frac{\overline{X}_o}{2} + \frac{(1 - \overline{X}_o)}{2} \frac{U(|b| - \cos\theta_c)}{1 - \cos\theta_c} \tag{18}$$

Taking the limit of strong association $\overline{X}_o \to 0$ and vanishing association site size $\theta_c \to 0$, the ODF simplifies to

$$\xi(b) = \frac{\delta(|b| - 1)}{2} \tag{19}$$

It is in this limit that CSW association sites reduce to the simpler spin up / down approach taken by Kierlik and Rosinberg.[46]

Now we turn our attention to the integral $I_A$ in Eq. (8). Due to the overlap function $O_{AB}$, this integral is only non-zero for orientations $|b_1| \geq \cos\theta_c$ which upon simplification yields the following

$$I_A(b_1) = f_{AB} \frac{\Psi}{2} \bar{\rho} U(|b_1| - \cos\theta_c) \int_{\cos\theta_c}^{1} db_2 \xi(b_2) X_B(b_2) \tag{20}$$

where $\Psi$ is the integral over the pair correlation function

$$\Psi = 2\int_{d}^{z_c} g(x)dx = \frac{2}{\bar{\rho}}\left(1 - \exp\left(-\frac{\bar{\rho}(z_c - d)}{(1 - \bar{\rho}d)}\right)\right) \tag{21}$$

We decompose the integral in Eq. (20) as

$$\int_{\cos\theta_c}^{1} db_2 \xi(b_2) X_B(b_2) = \int_{-1}^{1} db_2 \xi(b_2) X_B(b_2) - \int_{-1}^{-\cos\theta_c} db_2 \xi(b_2) X_B(b_2) - \int_{-\cos\theta_c}^{\cos\theta_c} db_2 \xi(b_2) X_B(b_2) \tag{22}$$

We evaluate the contributions on the right hand side of Eq. (22) as

$$\int_{-1}^{-\cos\theta_c} db_2 \xi(b_2) X_B(b_2) = \int_{\cos\theta_c}^{1} db_2 \xi(b_2) X_B(b_2)$$

$$\int_{-1}^{1} db_2 \xi(b_2) X_B(b_2) = \bar{X}_B \tag{23}$$

$$\int_{-\cos\theta_c}^{\cos\theta_c} db_2 \xi(b_2) X_B(b_2) = \bar{X}_o \cos\theta_c$$

Solving Eqns. (22) – (23)

$$\int_{\cos\theta_c}^{1} db_2 \xi(b_2) X_B(b_2) = \frac{1}{2}(\bar{X}_B - \bar{X}_o \cos\theta_c) = \frac{1}{2}\bar{X}_A\left(1 - \frac{\bar{X}_o}{\bar{X}_A}\cos\theta_c\right) \tag{24}$$

In Eq. (24) we have enforced the symmetry of association sites

$$\bar{X}_A = \bar{X}_B \tag{25}$$

Now Eqns. (7), (20) and (24) are combined to obtain

$$\frac{1}{X_A(b_1)} = 1 + f_{AB}\frac{\Psi}{2}\overline{\rho}\overline{X}_A \frac{\left(1 - \frac{\overline{X}_o}{\overline{X}_A}\cos\theta_c\right)}{2} U(|b_1| - \cos\theta_c) \quad (26)$$

Multiplying each side of Eq. (26) by $X_A(b_1)\xi(b_1)$, integrating over $b_1$ and employing Eqns. (24) – (25) the final form for the fraction $\overline{X}_A$ is obtained

$$1 = \overline{X}_A + f_{AB}\Psi\overline{\rho}\overline{X}_A^2 \kappa_{AB} \quad (27)$$

Where we have defined

$$\kappa_{AB} = \frac{\left(1 - \frac{\overline{X}_o}{\overline{X}_A}\cos\theta_c\right)^2}{4} \quad (28)$$

In Eq. (27) the last term on the right hand side gives the average fraction of spheres which are bonded at site $A$, $\overline{X}_{bA}$

$$\overline{X}_{bA} = f_{AB}\Psi\overline{\rho}\overline{X}_A^2 \kappa_{AB} \quad (29)$$

where $\overline{\rho}\Psi$ is the probability the two spheres are within bonding distance, $\overline{X}_A^2$ is the probability that both have an association site available to bond, and $\kappa_{AB}$ is the probability both spheres are oriented correctly for bond formation between site $A$ on sphere 1 and site $B$ on sphere 2.

It is $\kappa_{AB}$ which distinguishes the 1D association theory from its 3D counterpart. In 1D, the associating fluid is orientationally inhomogeneous which gives $\kappa_{AB}$ as a function of the degree of association. In 3D this is not the case[9] where $\kappa_{AB} = (1 - \cos\theta_c)^2/4$ is independent of the degree of association. In 1D, $\kappa_{AB}$ takes a minimum for $\overline{X}_o = \overline{X}_A = 1$ when there is little association and is a maximum for $\overline{X}_o \to 0$ in strongly associating systems. What this shows is that the probability that two molecules are oriented correctly to form an association bond is significantly enhanced if the two molecules exist in separate bonded clusters. For this case the orientation of both molecules is restricted such that $|b| \geq \cos\theta_c$ meaning the probability both or oriented correctly for bonding is ¼. Compare this to the case of two unbonded molecules (with $\theta_c = 20°$) which have probability of both being correctly oriented of 0.00909, and it is clear that the restricted orientations induced by association in 1D will enhance the degree of association in the system. This effect is captured faithfully in the new theory through Eq. (28).

Now we must determine the relation between the average monomer fraction $\overline{X}_o$ and $\overline{X}_A$. We begin with the relation (obtained from Eqns. (7) and (14))

$$\frac{1}{X_o(b_1)} = \frac{1}{X_A(b_1)X_B(b_1)} = 1 + I_A(b_1) + I_B(b_1) + I_A(b_1)I_B(b_1) \tag{30}$$

Multiplying by $X_o(b_1)\xi(b_1)$ and integrating we obtain

$$1 = \overline{X}_o + \int_{-1}^{1} I_A(b_1)X_o(b_1)\xi(b_1)db_1 + \int_{-1}^{1} I_B(b_1)X_o(b_1)\xi(b_1)db_1 + \int_{-1}^{1} I_A(b_1)I_B(b_1)X_o(b_1)\xi(b_1)db_1 \tag{31}$$

Where

$$\int_{-1}^{1} I_A(b_1)X_o(b_1)\xi(b_1)db_1 = \int_{-1}^{1} I_B(b_1)X_o(b_1)\xi(b_1)db_1 \tag{32}$$

$$= f_{AB}\Psi\overline{\rho}\overline{X}_A\sqrt{\kappa_{AB}} \int_{\cos\theta_c}^{1} X_o(b_1)\xi(b_1)db_1$$

Decomposing the integral on the right had side of Eq. (32) we obtain [in the same fashion as Eq. (22)]

$$\int_{\cos\theta_c}^{1} X_o(b_1)\xi(b_1)db = \overline{X}_o \frac{(1-\cos\theta_c)}{2} \tag{33}$$

Now focusing on the last term in Eq. (31) we obtain

$$\int_{-1}^{1} I_A(b_1)I_B(b_1)X_o(b_1)\xi(b_1)db_1 = \left(f_{AB}\Psi\overline{\rho}\overline{X}_A\right)^2 \kappa_{AB}\overline{X}_o \frac{(1-\cos\theta_c)}{4} \tag{34}$$

Combining Eqns. (31) – (34)

$$\frac{1}{\bar{X}_o} = 1 + f_{AB}\Psi\bar{\rho}\bar{X}_A\sqrt{\kappa_{AB}}(1-\cos\theta_c) + (f_{AB}\Psi\bar{\rho}\bar{X}_A)^2\kappa_{AB}\frac{(1-\cos\theta_c)}{4} \qquad (35)$$

Equation (35) can be simplified further by employing Eq. (27)

$$\frac{1}{\bar{X}_o} = \cos\theta_c + \frac{1-\cos\theta_c}{Y_A^2} \qquad (36)$$

Where in Eq. (36) we have defined $Y_A$ as the fraction of molecules which are not bonded at site $A$ and are in the allowed bonding orientations $|b_1| \geq \cos\theta_c$

$$\frac{1}{Y_A} = 1 + \frac{1}{2}\left(\frac{1}{\bar{X}_A}-1\right)\frac{1}{\sqrt{\kappa_{AB}}} \qquad (37)$$

In the Janus particle limit ($\theta_c = 90°$) all particles are in allowed bonding orientations and we obtain monomer fractions and ODF's similar to the 3D case where there is no preferred orientation $\xi(b) = 1/2$ and association at each site becomes independent $\bar{X}_o = Y_A^2$.

To obtain the average fractions $\bar{X}_o$ and $\bar{X}_A$, Eqns. (27) and (36) are solved with $\kappa_{AB}$ defined by Eq. (28).

Now that the bonding fractions and ODF have been calculated, the free energy Eq. (4) is simplified using Eqns. (7), (10) and (18)

$$\frac{A - A_{HS}^{EX}}{k_bTN} = \ln\bar{\rho}\Lambda + \cos\theta_c\bar{X}_o\ln\frac{\bar{X}_o}{2} - \bar{X}_A + (1-\bar{X}_o\cos\theta_c)\ln\left(\left[\frac{\bar{X}_o}{2}+\frac{1-\bar{X}_o}{2(1-\cos\theta_c)}\right]Y_A^2\right) \qquad (38)$$

To calculate the chemical potential $\mu$ we revert to the graphical formalism of Wertheim[5] to obtain

$$\frac{\mu(b) - \mu_{HS}^{EX}}{k_BT} = \ln\bar{\rho}\Lambda + \ln\xi(b) + 2\ln X_A(b) - \bar{\rho}(1-\bar{X}_A)\frac{\partial\ln\Psi}{\partial\bar{\rho}} \qquad (39)$$

As can be seen in Eq. (39), the excess chemical potential is a function of orientation. To obtain the averaged chemical potential we multiply each side of Eq. (39) by the ODF and integrate to obtain

$$\frac{\bar{\mu} - \mu_{HS}^{EX}}{k_B T} = \ln \bar{\rho}\Lambda + \cos\theta_c \bar{X}_o \ln \frac{\bar{X}_o}{2} + (1 - \bar{X}_o \cos\theta_c) \ln\left(\left[\frac{\bar{X}_o}{2} + \frac{1 - \bar{X}_o}{2(1 - \cos\theta_c)}\right] Y_A^2\right) - \bar{\rho}(1 - \bar{X}_A) \frac{\partial \ln \Psi}{\partial \bar{\rho}} \quad (40)$$

The final quantity to be calculated is the order parameter $S$

$$S = \frac{3 \overline{\cos^2 \theta} - 1}{2} \quad (41)$$

Where $\overline{\cos^2 \theta}$ which is obtained from Eqns. (13) and (18) as

$$\overline{\cos^2 \theta} = \frac{\bar{X}_o}{3} + \frac{(1 - \bar{X}_o)}{3} \frac{(1 - \cos^3 \theta_c)}{1 - \cos \theta_c} \quad (42)$$

Note the order parameter is described in terms of $\cos^2 \theta_c$, meaning it does not discriminate between left handed and right handed clusters. For systems in which all molecular orientation vectors lie parallel (or antiparallel) to the $z$ axis $S = 1$, while for systems with unperturbed ODF's $S = 0$. There are two interesting limits of Eq. (41). The first is the Janus particle limit $\theta_c = 90°$; for this case all possible orientations lead to $|b| \geq \cos\theta_c$, so there is no preferred orientations, resulting in $S = 0$. Alternatively, the small patch limit $\theta_c \rightarrow 0°$ at low temperature $T \rightarrow 0$ yields an order parameter $S = 1$. Figure 3 plots the order parameter versus $\theta_c$ for various degrees of association. As can be seen, decreasing monomer fraction and $\theta_c$ lead to an increased $S$.

This completes the development of the equation of state for two site associating molecules confined to a single spatial dimension, but with access to all orientation angles. To verify the accuracy of this approach we perform Monte Carlo simulations which are briefly discussed in section III.

## III. Monte Carlo simulations

To verify the accuracy of the theory we perform new *NVT* (constant number of molecules, volume, temperature) Monte Carlo simulations using the intermolecular potential in Eq. (1). The simulations were initialized by ordering spheres $1 - N$ along a single spatial dimension $z$ where each sphere $i$ only interacts with it's nearest neighbors $i - 1$ and $i + 1$. Periodic boundary conditions were employed. Trial moves consisted of an attempted displacement and reorientation of a sphere, with the standard[49] acceptance criteria. The maximum displacement of a sphere was chosen to be less than one diameter, so there was no possibility of particle order being disrupted. Simulations were allowed to equilibrate for $10^9$ trial moves and production runs of $10^9$ trial moves were used to generate ensemble averages. For all simulations we used $N = 500$ molecules.

## IV. Theory validation

To validate the new theory we compare theoretical predictions to the Monte Carlo simulation results discussed in III. We restrict our attention to a moderate density of $\rho^* = \bar{\rho}d = 0.5$ and a critical radius of $z_c = 1.2d$. In Fig. 4 we compare theory and simulation for the fraction of spheres bonded once $\bar{X}_1 = 2(\bar{X}_A - \bar{X}_o)$ and twice $\bar{X}_2 - 1 - \bar{X}_1 - \bar{X}_o$ for reduced association energies of $\varepsilon^* = \varepsilon_{AB}/k_BT = 3$, 5 and 8. The independent variable is the critical angle $\theta_c$ which various between the zero patch $\theta_c = 0°$ and Janus particle $\theta_c = 90°$ limits.

For small $\theta_c$ there is little association due to the large penalty in decreased orientational entropy associated with bond formation. As $\theta_c$ is increased, this penalty in orientational entropy is decreased and there is a sudden increase in both $\bar{X}_1$ and $\bar{X}_2$ for the case $\varepsilon^* = 8$, with the transition occurring more gradually for smaller $\varepsilon^*$. For $\varepsilon^* = 8$, $\bar{X}_2$ increases rapidly to a limiting value near 0.8, while the corresponding $\bar{X}_1$ for this case plateaus near 0.1. For comparison we also included predictions for $\bar{X}_2$ using a traditional TPT1 treatment[8] where the system is assumed orientationally homogeneous (e.g. assuming a constant ODF). As can be seen, this assumption leads to a drastic underprediction of the fraction of spheres bonded twice, while the new theory, which includes a non – constant ODF, is highly accurate. Interestingly, the two approaches agree in the Janus particle limit, this is a result of the fact that for Janus particles there will be no preferred orientation (see below Eq. 42).

Figure 5 compares theory and simulation for the order parameter $S$, at the same conditions discussed in Fig. 4. Focusing on the case $\varepsilon^* = 8$, we see $S = 0$ for very small $\theta_c$ where there is little association. At an angle $\theta_c \sim 5°$, there is a rapid increase in $S$ as $\theta_c$ is increased until the maximum $S \sim 0.86$ near $\theta_c = 18°$. Increasing $\theta_c$ further decreases $S$, until $S = 0$ is obtained at $\theta_c = 90°$. The observed maximum is a consequence of the fact that spheres with small association sites, small $\theta_c$, must be highly oriented to form an association bond. This results in strong order within a cluster, but also results in a large penalty in decreased orientational entropy. The location of the maximum in $S$ is the point where the association site is small enough such that molecules must be significantly oriented to associate, but not so small that association in the system is weak. Interestingly, the location of this maximum shifts to larger $\theta_c$ as association energy is decreased. Also observable from Fig. 5, is the fact that increasing $\varepsilon^*$ (or equivalently decreasing $T$) increases $S$ as a result of increased association. Theory and simulation are in excellent agreement.

It is known that molecules which interact with short range intermolecular potentials do not undergo a liquid / vapor phase transition.[50] While association may result in the formation of large associated clusters, these do not constitute a liquid phase. This can be seen in Fig. 6 which plots the reduced pressure $P^* = Pd/k_bT$ and order parameter $S$ against density for molecules with an association site size $\theta_c = 20°$ at the low temperature (high association energy) $\varepsilon^* = 15$. As can be seen through $S$, molecules transition from random orientations at very low densities to being locally oriented in clusters as density is increased. However, this does not constitute a phase transition as the pressure is a monotonic function of density, meaning the conditions for phase equilibria are not satisfied.

As a final test, we compare theory and simulation for the reduced excess internal energy $E^* = E^{EX}/Nk_BT$. In the theory we calculate $E^*$ from the average number of bonds as

$$E^* = -\varepsilon^* \left( \frac{\overline{X}_1}{2} + \overline{X}_2 \right) \tag{43}$$

Figure 7 gives the comparison of $E^*$ predicted from the new theory (solid curves), TPT1 (dashed curves) and simulation symbols. For $\varepsilon^* = 8$, there is a rapid decrease in energy near $\theta_c = 10°$ due to a sudden increase in the fraction of spheres being fully bonded (Fig. 4). At higher temperatures (lower $\varepsilon^*$) the decrease in energy is more gradual. The current theory faithfully matches the simulation results, while TPT1 significantly over predicts the internal energy of the system. This inaccuracy of TPT1 is the result of the underprediction of $\bar{X}_2$ as demonstrated in Fig. 4.

## V. Conclusions

We have developed a new theory for the self – assembly and equation of state of two site associating fluids in 1D pores. While restricted to a single spatial dimension, the associating species were allowed to explore all orientations. The choice of the conical square well association sites allows the description of the intermolecular potential in terms of a single orientation angle $\theta$. The orientation dependent intermolecular potential, in combination with the 1D spatial treatment, renders the system orientationally inhomogeneous. Once a molecule is bonded, it is forced to lie in a certain range of orientations. This decrease in orientation entropy significantly increases the probability molecules of forming multiple bonds. Ultimately this effect enhances association in the system. As this effect is not captured in first order perturbation theory, TPT1 fails to accurately describe this system. To test the theory, Monte Carlo simulations were performed and the theory was found to accurately predict the orientational structure of the fluid.

This work lays the foundation for numerous future studies. For instance, for the 2 site axisymmetric case considered here, the ODF depends on only a single orientation angle. However, if we instead considered a 4 site model we lose this axis symmetry and the ODF will become dependent on additional angles. The general approach taken in this paper would still be applicable, although the development will become more complicated. It is clear from this work, that associating fluids in low dimensions must be treated as orientationally inhomogeneous.

**Figure 1:**

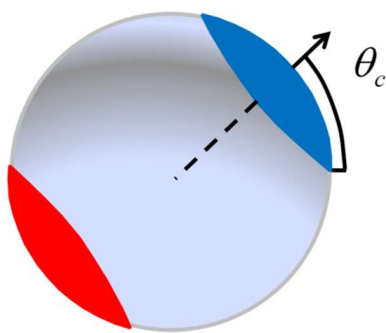

**Figure 1:** Diagram of two site (*A* – blue, *B* – red) associating sphere with CSW association sites

**Figure 2:**

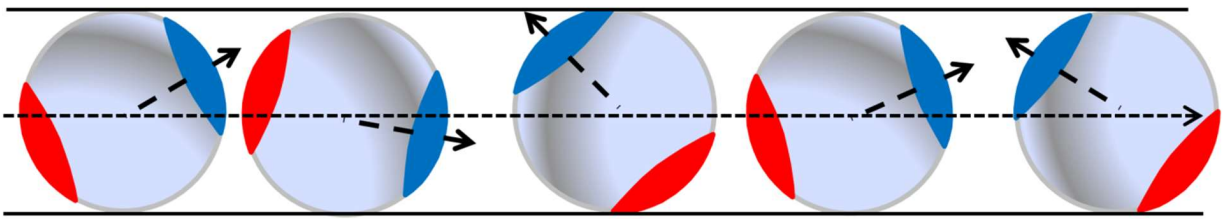

**Figure 2:** Diagram of 1-D fluid of associating spheres. Long dashed arrow represents *z* axis, and short dashed arrows through the *A* association sites define the orientation of spheres.

**Figure 3:**

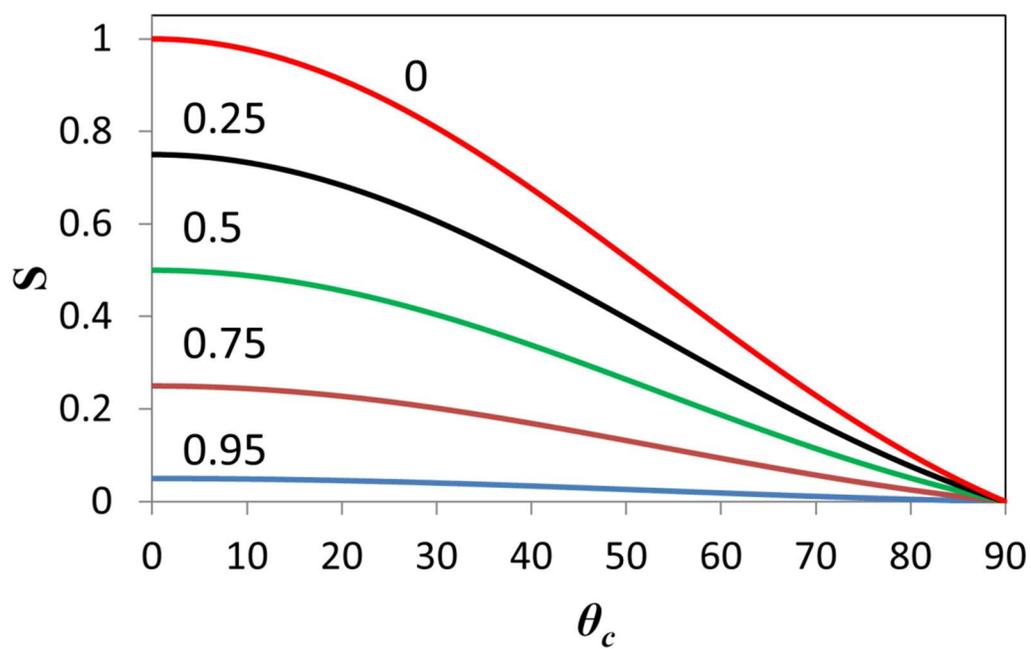

**Figure 3:** Order parameter (Eq. 41) versus $\theta_c$ (in degrees) for monomer fractions $\bar{X}_o$ = 0, 0.25, 0.5, 0.75 and 0.95

**Figure 4:**

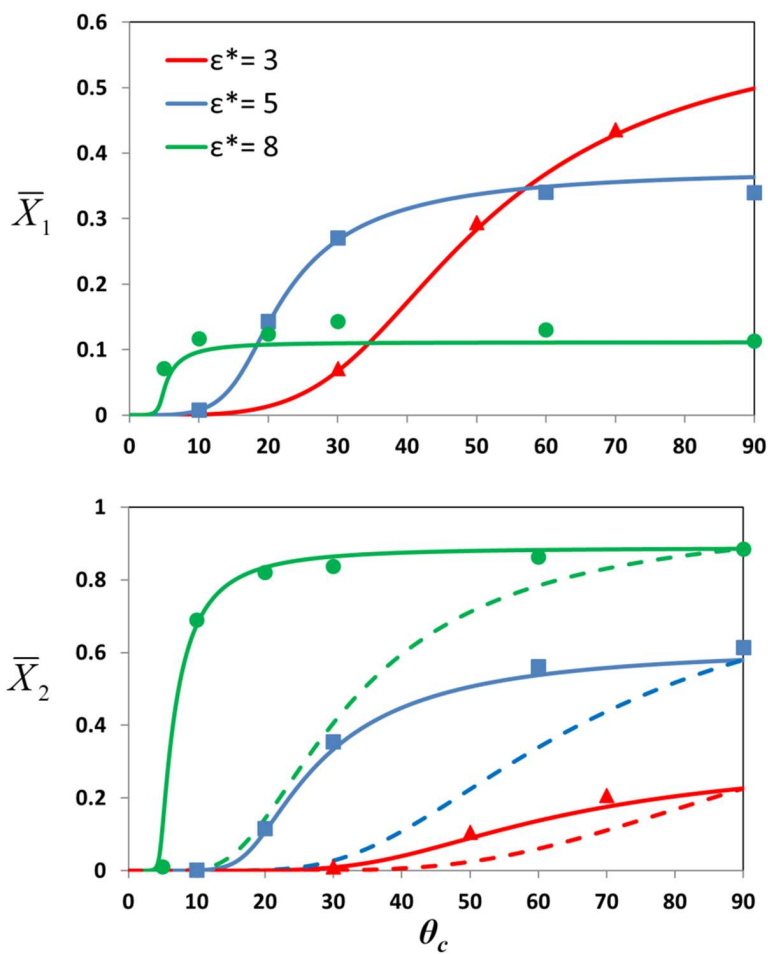

**Figure 4:** Fraction of molecules bonded once $\bar{X}_1$ (top) and twice $\bar{X}_2$ (bottom) versus $\theta_c$ (degrees) for reduced association energies $\varepsilon^* = 3$, 5 and 8. Curves give theory predictions and symbols are Monte Carlo simulation results triangle $\varepsilon^* = 3$, square $\varepsilon^* = 5$ and circle $\varepsilon^* = 8$. Solid curves give predictions of new theory, while dashed curves give predictions using the standard TPT1[8] treatment. Density is held constant at $\rho^* = 0.5$.

**Figure 5:**

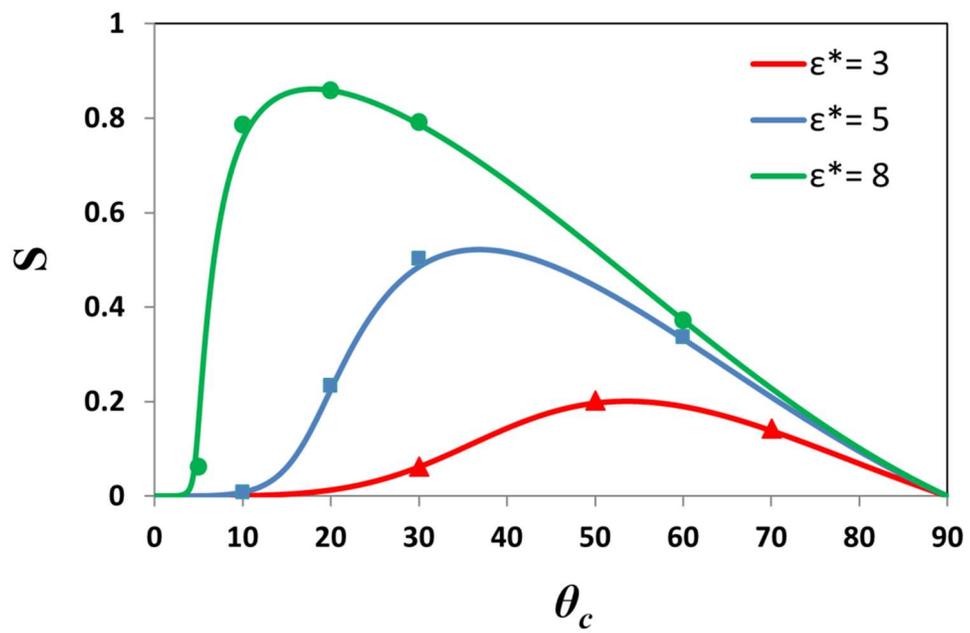

**Figure 5:** Same as figure 4, except calculations are for order parameter *S*.

**Figure 6:**

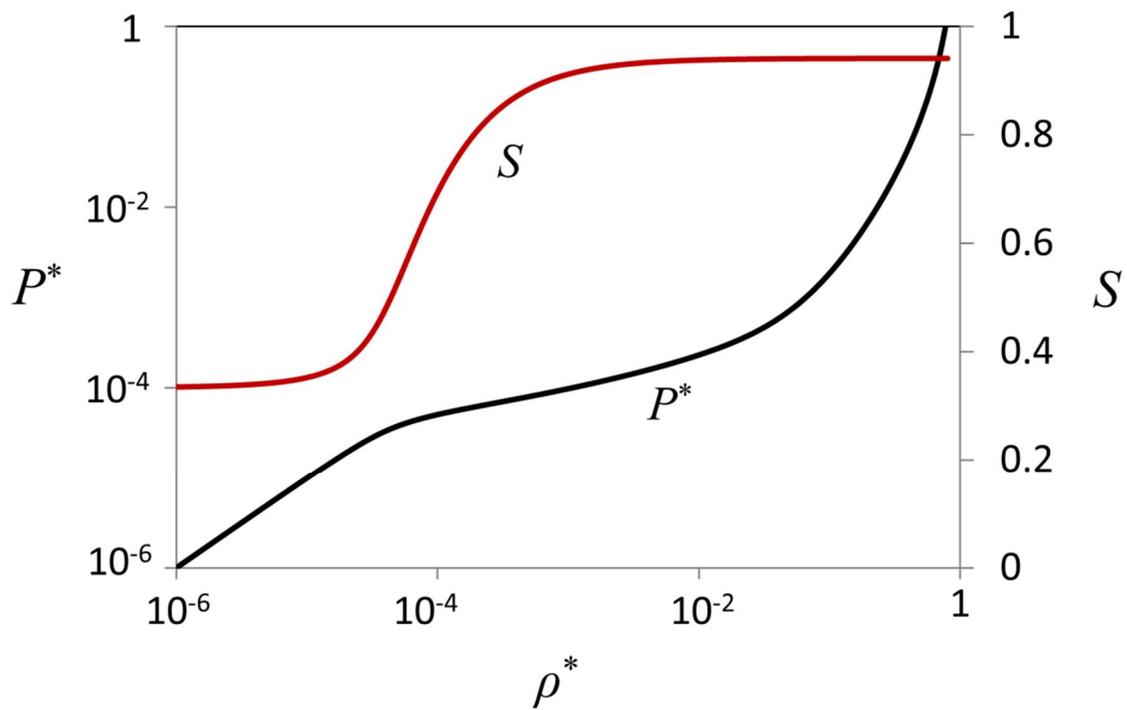

**Figure 6:** Reduced pressure (left axis) and order parameter (right axis) versus reduced density for the case $\theta_c = 20°$ and $\varepsilon^* = 15$.

**Figure 7:**

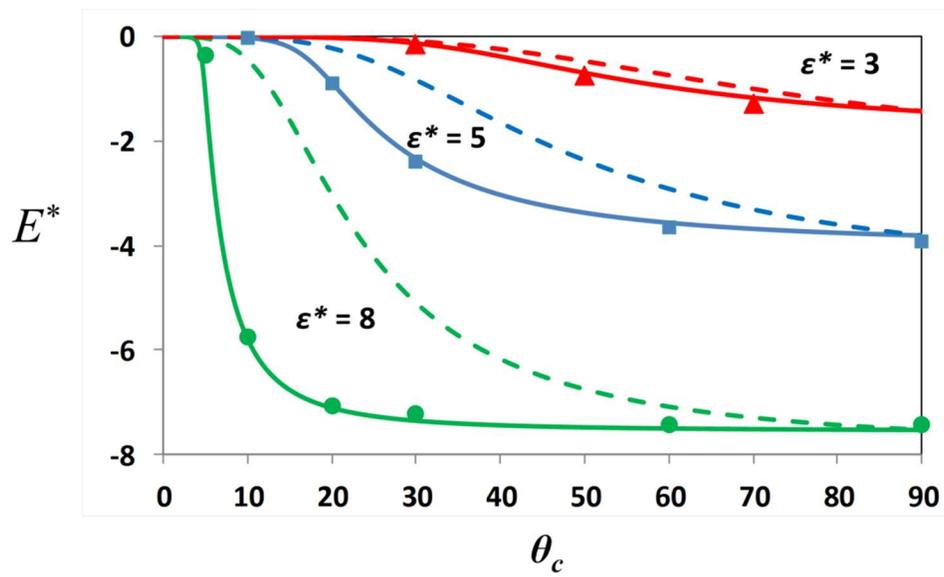

**Figure 7:** Same as Figure 4, except calculations are for reduced excess internal energy $E^*$